\documentclass[12pt,twocolumn,a4paper]{article}

\usepackage[export]{adjustbox}
\usepackage{amssymb,mathtools}
\usepackage[breaklinks]{hyperref}
\usepackage{cuted}
\usepackage[preprint]{neurips_2020}
\usepackage{hyperref}
\hypersetup{
    colorlinks=true,
    linkcolor=blue,
    filecolor=magenta,      
    urlcolor=blue,
    citecolor=blue
}

\usepackage[utf8]{inputenc} 
\usepackage[T1]{fontenc}    
\usepackage{enumitem}
\usepackage{url}            
\usepackage{graphicx}
\usepackage{subfigure}
\usepackage[skins]{tcolorbox}
\usepackage{hyperref}
\usepackage[T1]{fontenc}
\usepackage{epigraph}

\begin{document}

\begin{strip}%
 \centering
 \vskip 0.25in
{\LARGE\bf \fontfamily{cmr}\selectfont The Hardware Lottery} \\
\vspace{5mm}
{\fontfamily{cmr}\selectfont
\large Sara Hooker} \\
\vspace{5mm}
Google Research, Brain Team \\
 \vspace{1mm}
\texttt{shooker@google.com} \\
\vspace{5mm}
 {\large\bf \fontfamily{cmr}\selectfont Abstract} \\
 \vspace{5mm}
\begin{quote}%
{\fontfamily{cmr}\selectfont
Hardware, systems and algorithms research communities have
historically had different incentive structures and fluctuating
motivation to engage with each other explicitly. This historical
treatment is odd given that hardware and software have frequently
determined which research ideas succeed (and fail).
This essay introduces the term hardware lottery to describe when a research idea wins because it is suited to the available software and hardware and \emph{not} because the idea is superior to alternative research directions. Examples from early computer science history illustrate how hardware lotteries can delay research progress by casting successful ideas as failures. These lessons are particularly salient given the advent of domain specialized hardware which make it increasingly costly to stray off of the beaten path of research ideas. This essay posits that the gains from progress in computing are likely to become even more uneven, with certain research directions moving into the fast-lane while progress on others is further obstructed.
}
\par%
\end{quote}%
\vskip 1ex%
\end{strip}

\section{Introduction} \label{submission}
{\fontfamily{cmr}\selectfont History tells us that scientific progress is imperfect. Intellectual traditions and available tooling can prejudice scientists against certain ideas and towards others \citep{Kuhn1962}. This adds noise to the marketplace of ideas, and often means there is inertia in recognizing promising directions of research. In the field of artificial intelligence research, this essay posits that it is our tooling which has played a disproportionate role in deciding what ideas succeed (and which fail). 

What follows is part position paper and part historical review. This essay introduces the term \textit{hardware lottery} to describe when a research idea wins because it is compatible with available software and hardware and not because the idea is superior to alternative research directions. We argue that choices about software and hardware have often played a decisive role in deciding the winners and losers in early computer science history.}

\begin{figure*}[ht!]
\vskip 0.2in 
\begin{center}
  \includegraphics[width=0.8\textwidth]{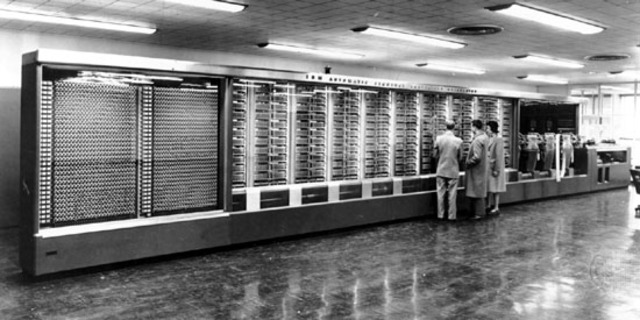}
    \caption{Early computers such as the Mark I were single use and were not expected to be re-purposed. While Mark I could be programmed to compute different calculations, it was essentially a very powerful calculator and could not run the variety of programs that we expect of our modern day machines.}
\label{fig:mark_1} \end{center} 
\end{figure*}

{\fontfamily{cmr}\selectfont
These lessons are particularly salient as we move into a new era of closer collaboration between hardware, software and machine learning research communities. After decades of treating hardware, software and algorithms as separate choices, the catalysts for closer collaboration include changing hardware economics \citep{Hennessy2019}, a “bigger is better” race in the size of deep learning architectures \citep{2018Amodei,2020arXiv200705558T} and the dizzying requirements of deploying machine learning to edge devices \citep{warden2019tinyml}.

Closer collaboration has centered on a wave of new generation hardware that is "domain specific" to optimize for commercial use cases of deep neural networks \citep{Jouppi2017,2019EdgeTpu,2020cortexm,Lee2018}. While domain specialization creates important efficiency gains for mainstream research focused on deep neural networks, it arguably makes it more even more costly to stray off of the beaten path of research ideas. An increasingly fragmented hardware landscape means that the gains from progress in computing will be increasingly uneven. While deep neural networks have clear commercial use cases, there are early warning signs that the path to the next breakthrough in AI may require an entirely different combination of algorithm, hardware and software. 

This essay begins by acknowledging a crucial paradox: machine learning researchers mostly ignore hardware despite the role it plays in determining what ideas succeed. In Section \ref{sect:seperate_tribes} we ask what has incentivized the development of software, hardware and machine learning research in isolation? Section \ref{sect:hardware_lottery} considers the ramifications of this siloed evolution with examples of early hardware and software lotteries. Today the hardware landscape is increasingly heterogeneous. This essay posits that the hardware lottery has not gone away, and the gap between the winners and losers will grow increasingly larger. Sections \ref{sect:persistance_hardware_lottery}-\ref{sect:likelyhood_future_hardware} unpack these arguments and Section \ref{sect:the_way_forward} concludes with some thoughts on what it will take to avoid future hardware lotteries.
}

\section{Separate Tribes} \label{sect:seperate_tribes}

\epigraph{It is not a bad description of man to describe him as a tool making animal.}{\textit{Charles Babbage, 1851}}

{\fontfamily{cmr}\selectfont
For the creators of the first computers the program was the machine. Early machines were single use and were not expected to be re-purposed for a new task because of both the cost of the electronics and a lack of cross-purpose software. Charles Babbage’s difference machine was intended solely to compute polynomial functions (1817)\citep{bruce1991}. Mark I  was a programmable calculator (1944)\citep{2014Mark1}. Rosenblatt’s perceptron machine computed a step-wise single layer network (1958)\citep{1986Rosenblatt}. Even the Jacquard loom, which is often thought of as one of the first programmable machines, in practice was so expensive to re-thread that it was typically threaded once to support a pre-fixed set of input fields (1804)\citep{posselt1888jacquard}.

The specialization of these early computers was out of necessity and not because computer architects thought one-off customized hardware was intrinsically better. However, it is worth pointing out that our own intelligence is both algorithm and machine. We do not inhabit multiple brains over the course of our lifetime. Instead, the notion of human intelligence is intrinsically associated with the physical 1400g of brain tissue and the patterns of connectivity between an estimated 85 billion neurons in your head \citep{2017neurons}.  When we talk about human intelligence, the prototypical image that probably surfaces as you read this is of a pink ridged cartoon blob. It is impossible to think of our cognitive intelligence without summoning up an image of the hardware it runs on.}

\begin{figure}[ht!]
\vskip 0.2in 
\begin{center}
  \includegraphics[width=0.8\linewidth]{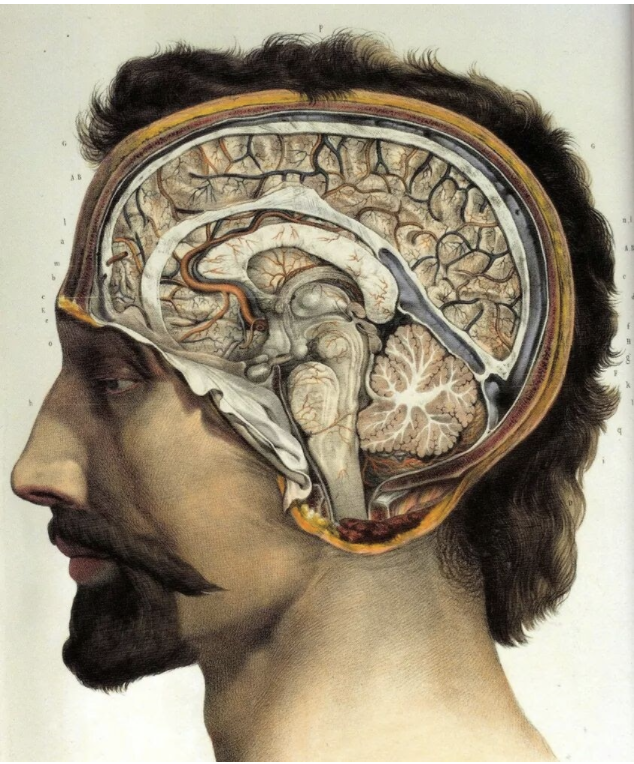}
    \caption{Our own cognitive intelligence is inextricably both hardware and algorithm. We do not inhabit multiple brains over our lifetime.}
\label{fig:human_brain} \end{center} 
\end{figure}

{\fontfamily{cmr}\selectfont
Today, in contrast to the necessary specialization in the very early days of computing, machine learning researchers tend to think of hardware, software and algorithm as three separate choices. This is largely due to a period in computer science history that radically changed the type of hardware that was made and incentivized hardware, software and machine learning research communities to evolve in isolation.}

\subsection{The General Purpose Era} \label{sect:general_purpose}

{\fontfamily{cmr}\selectfont
The general purpose computer era crystallized in 1969, when an opinion piece by a young engineer called Gordan Moore appeared in Electronics magazine with the apt title “Cramming more components onto circuit boards”\citep{moore1965}. Moore predicted you could double the amount of transistors on an integrated circuit every two years. Originally, the article and subsequent follow-up was motivated by a simple desire -- Moore thought it would sell more chips. However, the prediction held and motivated a remarkable decline in the cost of transforming energy into information over the next 50 years.

Moore’s law combined with Dennard scaling \citep{1050511} enabled a factor of three magnitude increase in microprocessor performance between 1980-2010 \citep{computerhistorymuseum}. The predictable increases in compute and memory every two years meant hardware design became risk-averse. Even for tasks which demanded higher performance, the benefits of moving to specialized hardware could be quickly eclipsed by the next generation of general purpose hardware with ever growing compute.

The emphasis shifted to universal processors which could solve a myriad of different tasks. Why experiment on more specialized hardware designs for an uncertain reward when Moore’s law allowed chip makers to lock in predictable profit margins? The few attempts to deviate and produce specialized supercomputers for research were financially unsustainable and short lived \citep{2018acceleratingai,1995thinkingmachines}. A few very narrow tasks like mastering chess were an exception to this rule because the prestige and visibility of beating a human adversary attracted corporate sponsorship \citep{Moravec98whenwill}.}

{\fontfamily{cmr}\selectfont
Treating the choice of hardware, software and algorithm as independent has persisted until recently. It is expensive to explore new types of hardware, both in terms of time and capital required. Producing a next generation chip typically costs \$30-80 million dollars and 2-3 years to develop \citep{2019Feldman}. These formidable barriers to entry have produced a hardware research culture that might feel odd or perhaps even slow to the average machine learning researcher. While the number of machine learning publications has grown exponentially in the last 30 years \citep{Dean202011TD}, the number of hardware publications have maintained a fairly even cadence \citep{singh_article}. For a hardware company, leakage of intellectual property can make or break the survival of the firm. This has led to a much more closely guarded research culture.

In the absence of any lever with which to influence hardware development, machine learning researchers rationally began to treat hardware as a sunk cost to work around rather than something fluid that could be shaped. However, just because we have abstracted away hardware does not mean it has ceased to exist. Early computer science history tells us there are many hardware lotteries where the choice of hardware and software has determined which ideas succeed (and which fail).
}

\begin{figure*}[ht]
\vskip 0.2in 
\begin{center}
  \includegraphics[width=0.8\textwidth]{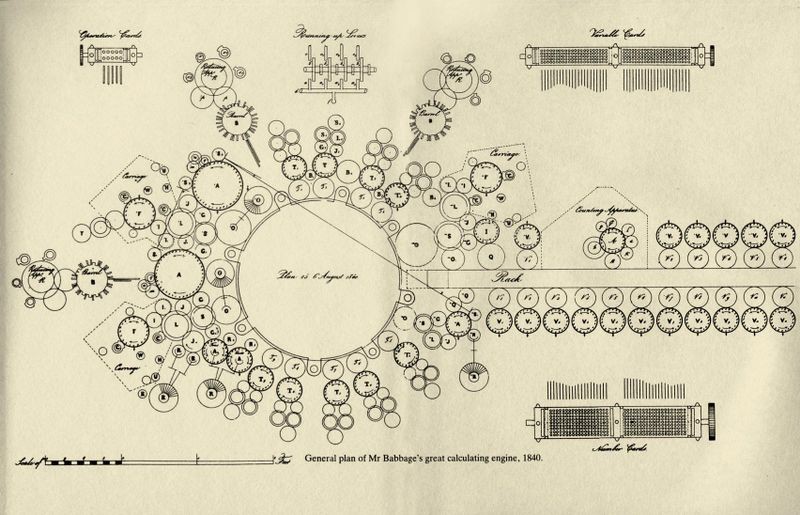}
    \caption{The analytical engine designed by Charles Babbage was never built in part because he had difficulty fabricating parts with the correct precision. This image depicts the general plan of the analytical machine in 1840.}
\label{fig:anayltical_machine} \end{center} 
\end{figure*}
    
\section{The Hardware Lottery} \label{sect:hardware_lottery}

\epigraph{I suppose it is tempting, if the only tool you have is a hammer, to treat everything as if it were a nail.}{\textit{Abraham Maslow}, 1966.}

{\fontfamily{cmr}\selectfont
The first sentence of Anna Karenina by Tolstoy reads “Happy families are all alike, every unhappy family is unhappy in it’s own way.” \citep{tolstoy2016anna}. Tolstoy is saying that it takes many different things for a marriage to be happy -- financial stability, chemistry, shared values, healthy offspring. However, it only takes one of these aspects to not be present for a family to be unhappy. This has been popularized as the Anna Karenina principle -- “a deficiency in any one of a number of factors dooms an endeavor to failure.” \citep{dwayne2001}.

Despite our preference to believe algorithms succeed or fail in isolation, history tells us that most computer science breakthroughs follow the Anna Karenina principle. Successful breakthroughs are often distinguished from failures by benefiting from multiple criteria aligning serendipitously. For AI research, this often depends upon winning what this essay terms the \textit{hardware lottery} — avoiding possible points of failure in downstream hardware and software choices.

An early example of a hardware lottery is the analytical machine (1837). Charles Babbage was a computer pioneer who designed a machine that (at least in theory) could be programmed to solve any type of computation. His analytical engine was never built in part because he had difficulty fabricating parts with the correct precision \citep{Raymond1990}. The electromagnetic technology to actually build the theoretical foundations laid down by Babbage only surfaced during WWII. In the first part of the 20th century, electronic vacuum tubes were heavily used for radio communication and radar. During WWII, these vacuum tubes were re-purposed to provide the compute power necessary to break the German enigma code \citep{2018vacuum}.

As noted in the TV show Silicon Valley, often “being too early is the same as being wrong.” When Babbage passed away in 1871, there was no continuous path between his ideas and modern day computing. The concept of a stored program, modifiable code, memory and conditional branching were rediscovered a century later because the right tools existed to empirically show that the idea worked.
}

\subsection{The Lost Decades} \label{sect:lost_decades}

{\fontfamily{cmr}\selectfont
Perhaps the most salient example of the damage caused by not winning the hardware lottery is the delayed recognition of deep neural networks as a promising direction of research. Most of the algorithmic components to make deep neural networks work had already been in place for a few decades: backpropagation (invented in 1963 \citep{1963steinbuch}, reinvented in 1976 \citep{Linnainmaa1976TaylorEO}, and again in 1988 \citep{1988rumelhart}), deep convolutional neural networks (\citep{FUKUSHIMA1982455}, paired with backpropagation in 1989 \citep{LeCun1989}). However, it was only three decades later that deep neural networks were widely accepted as a promising research direction.

This gap between algorithmic advances and empirical success is in large part due to incompatible hardware. During the general purpose computing era, hardware like CPUs were heavily favored and widely available. CPUs are very good at executing any set of complex instructions but incur high memory costs because of the need to cache intermediate results and process one instruction at a time \citep{2018Sato}. This is known as the von Neumann Bottleneck — the available compute is restricted by “the lone channel between the CPU and memory along which data has to travel sequentially” \citep{1985understandingcomputers}.

The von Neumann bottleneck was terribly ill-suited to matrix
multiplies, a core component of deep neural network architectures. Thus, training on CPUs quickly exhausted memory bandwidth and it simply wasn’t possible to train deep neural networks with multiple layers. The need for hardware that supported tasks with lots of parallelism was pointed out as far back as the early 1980s in a series of essays titled “Parallel Models of Associative Memory” \citep{Hinton1989}. The essays argued persuasively that biological evidence suggested massive parallelism was needed to make deep neural network approaches work \citep{rumelhart1987}. 

In the late 1980/90s, the idea of specialized hardware for neural networks had passed the novelty stage \citep{MISRA2010239,Lindsey1994,jeff1990}. However, efforts remained fractured by lack of shared software and the cost of hardware development. Most of the attempts that were actually operationalized like the Connection Machine in 1985 \citep{1995thinkingmachines}, Space in 1992 \citep{Howe1994}, Ring Array Processor in 1989 \citep{MORGAN1992248} and the Japanese 5th generation computer project \citep{morgan1983} were designed to favor logic programming such as PROLOG and LISP that were poorly suited to connectionist deep neural networks. Later iterations such as HipNet-1 \citep{kingsburyhipnet}, and the Analog Neural Network Chip in 1991 \citep{Sackinger129422} were promising but short lived because of the intolerable cost of iteration and the need for custom silicon. Without a consumer market, there was simply not the critical mass of end users to be financially viable.
}

\begin{figure}[ht!]
\vskip 0.2in 
\begin{center}
  \includegraphics[width=0.9\linewidth]{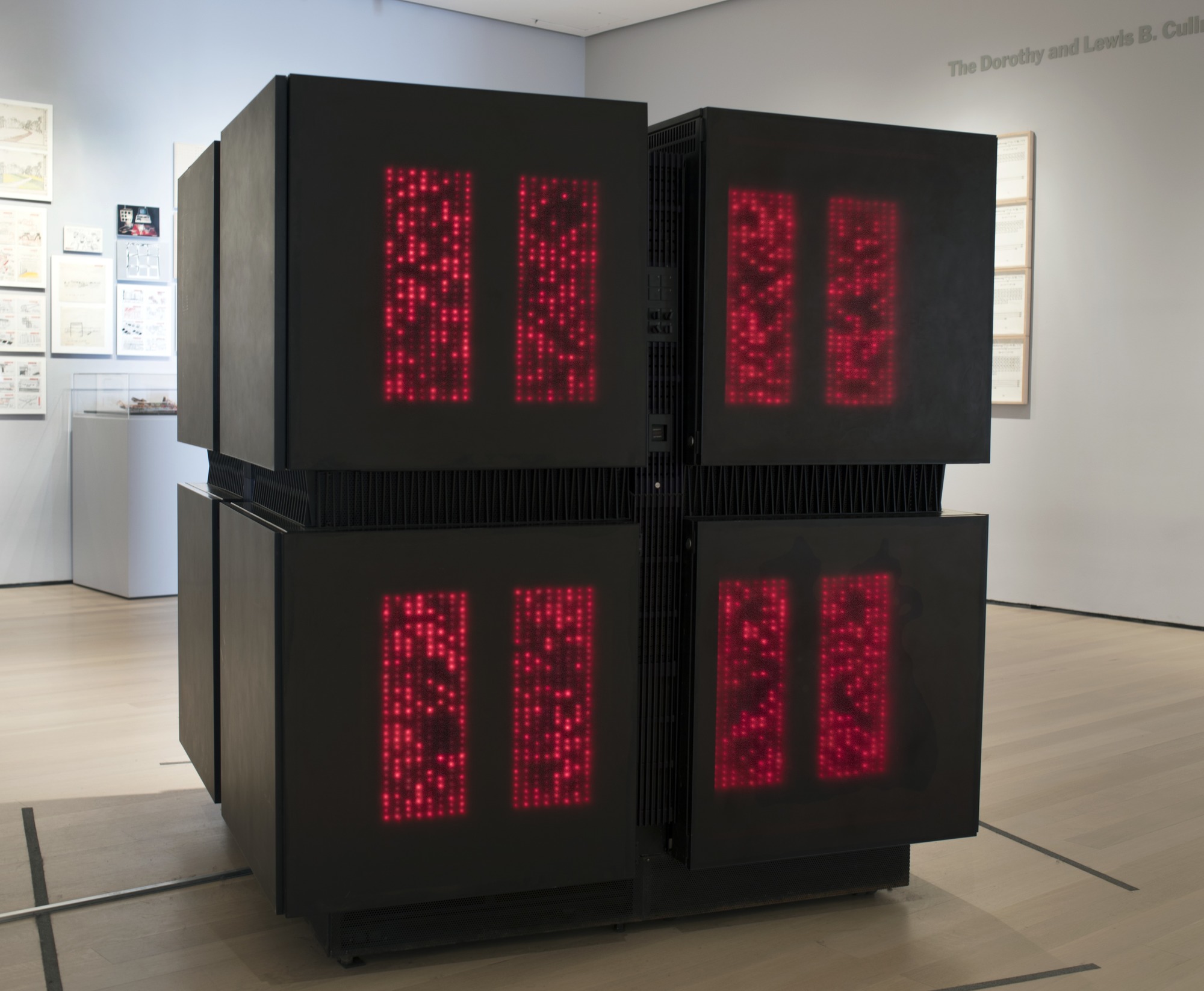}
    \caption{The connection machine was one of the few examples of hardware that deviated from general purpose cpus in the 1980s/90s. Thinking Machines ultimately went bankrupt after the inital funding from DARPA dried up.}
\label{fig:connection_machine} \end{center} 
\end{figure}

{\fontfamily{cmr}\selectfont
It would take a hardware fluke in the early 2000s, a full four decades after the first paper about backpropagation was published, for the insight about massive parallelism to be operationalized in a useful way for connectionist deep neural networks. Many inventions are re-purposed for means unintended by their designers. Edison’s phonograph was never intended to play music. He envisioned it as preserving the last words of dying people or teaching spelling. In fact, he was disappointed by its use playing popular music as he thought this was too “base” an application of his invention \citep{diamond98}. In a similar vein, deep neural networks only began to work when an existing technology was unexpectedly re-purposed.

A graphical processing unit (GPU) was originally introduced in the 1970s as a specialized accelerator for video games and for developing graphics for movies and animation. In the 2000s, like Edison’s phonograph, GPUs were re-purposed for an entirely unimagined use case -- to train deep neural networks \citep{Chellapilla2006,OH20041311kyoung,claudiu2010,Fatahalian2004,Payne2005}. GPUs had one critical advantage over CPUs - they were far better at parallelizing a set of simple decomposable instructions such as matrix multiples \citep{BRODTKORB20134,Dettmers2020}. This higher number of floating operation points per second (FLOPS) combined with clever distribution of training between GPUs unblocked the training of deeper networks. The number of layers in a network turned out to be the key. Performance on ImageNet jumped with ever deeper networks in 2011 \citep{ciresan2011}, 2012 \citep{NIPS2012_4824} and 2015 \citep{7298594}. A striking example of this jump in efficiency is a comparison of the now famous 2012 Google paper which used 16,000 CPU cores to classify cats \citep{quoc2012} to a paper published a mere year later that solved the same task with only two CPU cores and four GPUs \citep{coates13}.
}

\begin{figure}[ht!]
\vskip 0.2in 
\begin{center}
  \includegraphics[width=0.8\linewidth]{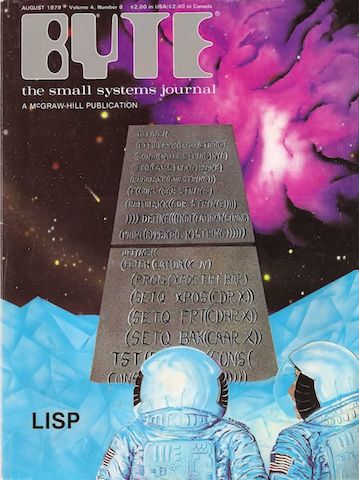}
    \caption{Byte magazine cover, August 1979, volume 4. LISP was the dominant language for artificial intelligence research through the 1990's. LISP was particularly well suited to handling logic expressions, which were a core component of reasoning and expert systems.}
\label{fig:lisp_prolog} \end{center} 
\end{figure}

\subsection{Software Lotteries} \label{sect:software_lotteries}

{\fontfamily{cmr}\selectfont
Software also plays a role in deciding which research ideas win and lose. Prolog and LISP were two languages heavily favored until the mid-90’s in the AI community. For most of this period, students of AI were expected to actively master one or both of these languages \citep{lucas1991}. LISP and Prolog were particularly well suited to handling logic expressions, which were a core component of reasoning and expert systems. 

For researchers who wanted to work on connectionist ideas like deep neural networks there was not a clearly suited language of choice until the emergence of Matlab in 1992 \citep{Demuth93neuralnetwork}. Implementing connectionist networks in LISP or Prolog was cumbersome and most researchers worked in low level languages like c++ \citep{lispcode}. It was only in the 2000’s that there started to be a more healthy ecosystem around software developed for deep neural network approaches with the emergence of LUSH \citep{lush2002} and subsequently TORCH \citep{Torch2002}. 

Where there is a loser, there is also a winner. From the 1960s through the mid 80s, most mainstream research was focused on symbolic approaches to AI \citep{Haugeland}. Unlike deep neural networks where learning an adequate representation is delegated to the model itself, symbolic approaches aimed to build up a knowledge base and use decision rules to replicate how humans would approach a problem. This was often codified as a sequence of logic what-if statements that were well suited to LISP and PROLOG. The widespread and sustained popularity of symbolic approaches to AI cannot easily be seen as independent of how readily it fit into existing programming and hardware frameworks.
}

\section{The Persistence of the Hardware Lottery} \label{sect:persistance_hardware_lottery}

{\fontfamily{cmr}\selectfont
Today, there is renewed interest in joint collaboration between hardware, software and machine learning communities. We are experiencing a second pendulum swing back to specialized hardware. The catalysts include changing hardware economics prompted by the end of Moore’s law and the breakdown of Dennard scaling \citep{Hennessy2019}, a “bigger is better” race in the number of model parameters that has gripped the field of machine learning \citep{2018Amodei}, spiralling energy costs \citep{2014Horowitz,strubell2019energy} and the dizzying requirements of deploying machine learning to edge devices \citep{warden2019tinyml}.

The end of Moore’s law means we are not guaranteed more compute, hardware will have to earn it. To improve efficiency, there is a shift from task agnostic hardware like CPUs to domain specialized hardware that tailor the design to make certain tasks more efficient. The first examples of domain specialized hardware released over the last few years -- TPUs \citep{Jouppi2017}, edge-TPUs \citep{2019EdgeTpu}, Arm Cortex-M55 \citep{2020cortexm}, Facebook's \textit{big sur} \citep{Lee2018} -- optimize explicitly for costly operations common to deep neural networks like matrix multiplies.

Closer collaboration between hardware and research communities will undoubtedly continue to make the training and deployment of deep neural networks more efficient. For example, unstructured pruning \citep{hooker2019,gale2019state,2019arXiv191111134E} and weight specific quantization \citep{Dong2019} are very successful compression techniques in deep neural networks but are incompatible with current hardware and compilation kernels. 

While these compression techniques are currently not supported, many clever hardware architects are currently thinking about how to solve for this. It is a reasonable prediction that the next few generations of chips or specialized kernels will correct for hardware biases against these techniques \citep{Wang2018HAQHA,2020sun}. Some of the first designs to facilitate sparsity have already hit the market \citep{nvidia2020}. In parallel, there is interesting research developing specialized software kernels to support unstructured sparsity \citep{Elsen_2020_CVPR,2020arXiv200610901G,Gray2017GPUKF}.

In many ways, hardware is catching up to the present state of machine learning research. Hardware is only economically viable if the lifetime of the use case lasts more than three years \citep{Dean202011TD}. Betting on ideas which have longevity is a key consideration for hardware developers. Thus, co-design effort has focused almost entirely on optimizing an older generation of models with known commercial use cases. For example, matrix multiplies are a safe target to optimize for because they are here to stay — anchored by the widespread use and adoption of deep neural networks in production systems. Allowing for unstructured sparsity and weight specific quantization are also safe targets because there is wide consensus that these will enable higher levels of compression.  

There is still a separate question of whether hardware innovation is versatile enough to unlock or keep pace with entirely new machine learning research directions. It is difficult to answer this question because data points here are limited -- it is hard to model the counterfactual of would this idea succeed given different hardware. However, despite the inherent challenge of this task, there is already compelling evidence that domain specialized hardware makes it more costly for research ideas that stray outside of the mainstream to succeed.

In 2019, a paper was published called “Machine learning is stuck in a rut.” \citep{Barham2019}. The authors consider the difficulty of training a new type of computer vision architecture called capsule networks \citep{NIPS2017_6975} on domain specialized hardware. Capsule networks include novel components like squashing operations and routing by agreement. These architecture choices aimed to solve for key deficiencies in convolutional neural networks (lack of rotational invariance and spatial hierarchy understanding) but strayed from the typical architecture of neural networks. As a result, while capsule networks operations can be implemented reasonably well on CPUs, performance falls off a cliff on accelerators like GPUs and TPUs which have been overly optimized for matrix multiplies.

Whether or not you agree that capsule networks are the future of computer vision, the authors say something interesting about the difficulty of trying to train a new type of image classification architecture on domain specialized hardware. Hardware design has prioritized delivering on commercial use cases, while built-in flexibility to accommodate the next generation of research ideas remains a distant secondary consideration.

While specialization makes deep neural networks more efficient, it also makes it far more costly to stray from accepted building blocks. It prompts the question of how much researchers will implicitly overfit to ideas that operationalize well on available hardware rather than take a risk on ideas that are not currently feasible? What are the failures we still don’t have the hardware and software to see as a success?
}

\section{The Likelyhood of Future Hardware Lotteries} \label{sect:likelyhood_future_hardware}

\epigraph{What we have before us are some breathtaking opportunities disguised as insoluble problems.}{\textit{ John Gardner, 1965.}}

{\fontfamily{cmr}\selectfont
It is an ongoing, open debate within the machine learning community about how much future algorithms will differ from models like deep neural networks \citep{thebitterlesson2019,welling2019}. The risk you attach to depending on domain specialized hardware is tied to your position on this debate. Betting heavily on specialized hardware makes sense if you think that future breakthroughs depend upon pairing deep neural networks with ever increasing amounts of data and computation.

Several major research labs are making this bet, engaging in a “bigger is better” race in the number of model parameters and collecting ever more expansive datasets. However, it is unclear whether this is sustainable. An algorithms scalability is often thought of as the performance gradient relative to the available resources. Given more resources, how does performance increase?

For many subfields, we are now in a regime where the rate of return for additional parameters is decreasing \citep{2020Thompson,2020brown}. For example, while the parameters almost double between Inception V3 \citep{2016Szegedy}and Inception V4 architectures \citep{2015szegedy} (from $21.8$ to $41.1$ million parameters), accuracy on ImageNet differs by less than 2\% between the two networks (78.8 vs 80 \%) \citep{kornblith2018}. The cost of throwing additional parameters at a problem is becoming painfully obvious. The training of GPT-3 alone is estimated to exceed \$12 million dollars \citep{wigger2020}. 

Perhaps more troubling is how far away we are from the type of intelligence humans demonstrate. Human brains despite their complexity remain extremely energy efficient. Our brain has over 85 billion neurons but runs on the energy equivalent of an electric shaver \citep{2017neurons}. While deep neural networks may be scalable, it may be prohibitively expensive to do so in a regime of comparable intelligence to humans. An apt metaphor is that we appear to be trying to build a ladder to the moon.}

\begin{figure}[ht]
\begin{center}
  \includegraphics[width=0.8\linewidth]{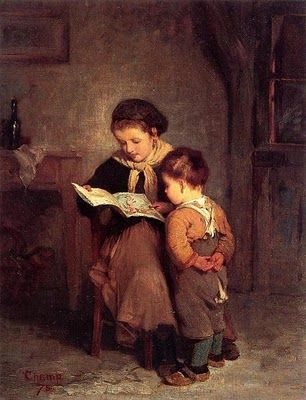}
    \caption{Human latency for certain tasks suggests we have specialized pathways for different stimuli. For example, it is easy for a human to walk and talk at the same time. However, it is far more cognitively taxing to attempt to read and talk.}
\label{fig:reading_walking} \end{center} 
\end{figure}

{\fontfamily{cmr}\selectfont
Biological examples of intelligence differ from deep neural networks in enough ways to suggest it is a risky bet to say that deep neural networks are the only way forward. While algorithms like deep neural networks rely on global updates in order to learn a useful representation, our brains do not. Our own intelligence relies on decentralized local updates which surface a global signal in ways that are still not well understood \citep{LILLICRAP201982,Marblestone2016,Bi10464}. 

In addition, our brains are able to learn efficient representations from far fewer labelled examples than deep neural networks \citep{Zador2019ACO}. For typical deep learning models the entire model is activated for every example which leads to a quadratic blow-up in training cost. In contrast, evidence suggests that the brain does not perform a full forward and backward pass for all inputs. Instead, the brain simulates what inputs are expected against incoming sensory data. Based upon the certainty of the match, the brain simply infills. What we see is largely virtual reality computed from memory \citep{Eagleman2036,bubic2010,Heeger1773}. 

Humans have highly optimized and specific pathways developed in our biological hardware for different tasks \citep{von2000computer,marcus2014,Kennedy2000SignalprocessingMA}. For example, it is easy for a human to walk and talk at the same time. However, it is far more cognitively taxing to attempt to read and talk \citep{Stroop1935}. This suggests the way a network is organized and our inductive biases is as important as the overall size of the network \citep{HerculanoHouzel2014TheEB,inductive_biases,spelke2007}. Our brains are able to fine-tune and retain skills across our lifetime \citep{benna2016,bremner2013,stein2004handbook,tani2016exploring,Gallistel2009,Tulving2002,Barnett2002WhenAW}. In contrast, deep neural networks that are trained upon new data often evidence catastrophic forgetting, where performance deteriorates on the original task because the new information interferes with previously learned behavior \citep{Mcclelland1995,MCCLOSKEY1989109,2018Parisi}.

The point of these examples is not to convince you that deep neural networks are not the way forward. But, rather that there are clearly other models of intelligence which suggest it may not be the only way. It is possible that the next breakthrough will require a fundamentally different way of modelling the world with a different combination of hardware, software and algorithm. We may very well be in the midst of a present day hardware lottery.
}

\section{The Way Forward} \label{sect:the_way_forward}
\epigraph{Any machine coding system should be judged quite largely from the point of view of how easy it is for the operator to obtain results.}{\textit{John Mauchly, 1973.}}

{\fontfamily{cmr}\selectfont
Scientific progress occurs when there is a confluence of factors which allows the scientist to overcome the "stickyness" of the existing paradigm. The speed at which paradigm shifts have happened in AI research have been disproportionately determined by the degree of alignment between hardware, software and algorithm. Thus, any attempt to avoid hardware lotteries must be concerned with making it cheaper and less time-consuming to explore different hardware-software-algorithm combinations.

This is easier said than done. Expanding the search space of possible hardware-software-algorithm combinations is a formidable goal. It is expensive to explore new types of hardware, both in terms of time and capital required. Producing a next generation chip typically costs \$30-80 million dollars and takes $2$-$3$ years to develop \citep{2019Feldman}. The fixed costs alone of building a manufacturing plant are enormous; estimated at \$7 billion dollars in 2017 \citep{Thompson2018TheDO}. 

Experiments using reinforcement learning to optimize chip placement may help decrease cost \citep{2020Mirhoseini}. There is also renewed interest in re-configurable hardware such as field program gate array (FPGAs) \citep{Hauck2007} and coarse-grained reconfigurable arrays (CGRAs) \citep{8192487}. These devices allow the chip logic to be re-configured to avoid being locked into a single use case. However, the trade-off for flexibility is far higher FLOPS and the need for tailored software development. Coding even simple algorithms on FPGAs remains very painful and time consuming \citep{Shalf2020TheFO}.  

In the short to medium term hardware development is likely to remain expensive. The cost of producing hardware is important because it determines the amount of risk and experimentation hardware developers are willing to tolerate. Investment in hardware tailored to deep neural networks is assured because neural networks are a cornerstone of enough commercial use cases. The widespread profitability of deep learning has spurred a healthy ecosystem of hardware startups that aim to further accelerate deep neural networks \citep{Cade2018} and has encouraged large companies to develop custom hardware in-house \citep{7866802,Jouppi2017,Lee2018}.

The bottleneck will continue to be funding hardware for use cases that are not immediately commercially viable. These more risky directions include biological hardware \citep{tan2007,macia2014,Kriegman1853}, analog hardware with in-memory computation \citep{ambrogio2018}, neuromorphic computing \citep{8741810}, optical computing \citep{Lin1004}, and quantum computing based approaches \citep{2019cross}. There are also high risk efforts to explore the development of transistors using new materials \citep{Colwell2013,6527325}.}

\begin{figure}[ht!]
\vskip 0.2in 
\begin{center}
  \includegraphics[width=0.8\linewidth]{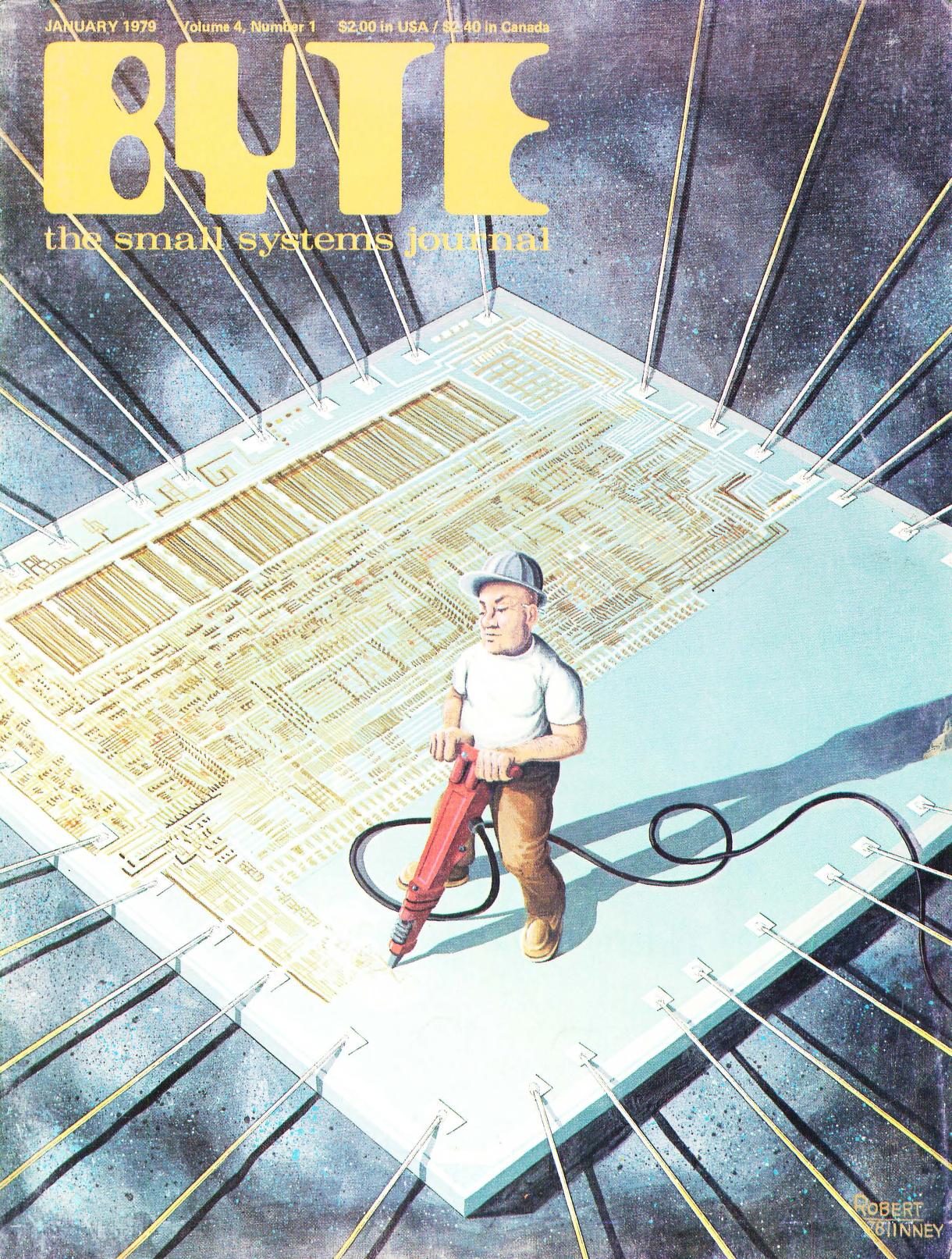}
    \caption{Byte magazine cover, March 1979, volume 4. Hardware design remains risk adverse due to the large amount of capital and time required to fabricate each new generation of hardware.}
\label{fig:hardware_bytes} \end{center} 
\end{figure}

{\fontfamily{cmr}\selectfont
Lessons from previous hardware lotteries suggest that investment must be sustained and come from both private and public funding programs. There is a slow awakening of public interest in providing such dedicated resources, such as the 2018 DARPA Electronics Resurgence Initiative which has committed to $1.5$ billion dollars in funding for microelectronic technology research \citep{DARPA018}. China has also announced a $47$ billion dollar fund to support semiconductor research \citep{Kubota2018}. However, even investment of this magnitude may still be woefully inadequate, as hardware based on new materials requires long lead times of $10$-$20$ years and public investment is currently far below industry levels of R\&D \citep{Shalf2020TheFO}.}

\subsection{A Software Revolution}\label{sect:software_revolution}
{\fontfamily{cmr}\selectfont
An interim goal should be to provide better feedback loops to researchers about how our algorithms interact with the hardware we do have. Machine learning researchers do not spend much time talking about how hardware chooses which ideas succeed and which fail. This is primarily because it is hard to quantify the cost of being concerned. At present, there are no easy and cheap to use interfaces to benchmark algorithm performance against multiple types of hardware at once. There are frustrating differences in the subset of software operations supported on different types of hardware which prevent the portability of algorithms across hardware types \citep{HotelSoftwarePF}. Software kernels are often overly optimized for a specific type of hardware which causes large discrepencies in efficiency when used with different hardware \citep{Hennessy2019}. 

These challenges are compounded by an ever more formidable and heterogeneous hardware landscape \citep{Reddi2020,7459430}. As the hardware landscape becomes increasingly fragmented and specialized, fast and efficient code will require ever more niche and specialized skills to write \citep{Lee2011}. This means that there will be increasingly uneven gains from progress in computer science research. While some types of hardware will benefit from a healthy software ecosystem, progress on other languages will be sporadic and often stymied by a lack of critical end users \citep{Thompson2018TheDO,Leisersoneaam9744}. 

One way to mitigate this need for specialized software expertise is to focus on the development of domain-specific languages which cater to a narrow domain. While you give up expressive power, domain-specific languages permit greater portability across different types of hardware. It allows developers to focus on the intent of the code without worrying about implementation details. \citep{Olukotun2014,Mernik2005,Cong2011}. Another promising direction is automatically auto-tuning the algorithmic parameters of a program based upon the downstream choice of hardware. This facilitates easier deployment by tailoring the program to achieve good performance and load balancing on a variety of hardware \citep{8476161,CLINTWHALEY20013,Asanovic2006,Ansel2014}.

The difficulty of both these approaches is that if successful, this further abstracts humans from the details of the implementation. In parallel, we need better profiling tools to allow researchers to have a more informed opinion about how hardware and software should evolve. Ideally, software could even surface recommendations about what type of hardware to use given the configuration of an algorithm. Registering what differs from our expectations remains a key catalyst in driving new scientific discoveries. 

Software needs to do more work, but it is also well positioned to do so. We have neglected efficient software throughout the era of Moore's law, trusting that predictable gains in compute would compensate for inefficiencies in the software stack. This means there are many low hanging fruit as we begin to optimize for more efficient code \citep{larus2008spending,Xu2010}. 
}
\vskip -1in 
\section{Conclusion}\label{sect:conclusion}
\vskip -0.1in 
{\fontfamily{cmr}\selectfont
George Gilder, an American investor, described the computer chip as “inscribing worlds on grains of sand” \citep{gilder2000telecosm}. The performance of an algorithm is fundamentally intertwined with the hardware and software it runs on. This essay proposes the term hardware lottery to describe how these downstream choices determine whether a research idea succeeds or fails. Today the hardware landscape is increasingly heterogeneous. This essay posits that the hardware lottery has not gone away, and the gap between the winners and losers will grow increasingly larger. In order to avoid future hardware lotteries, we need to make it easier to quantify the opportunity cost of settling for the hardware and software we have.
}

\section{Acknowledgments}
{\fontfamily{cmr}\selectfont
Thank you to many of my wonderful colleagues and peers who took time to provide valuable feedback on earlier versions of this essay. In particular, I would like to acknowledge the valuable input of Utku Evci, Erich Elsen, Melissa Fabros, Amanda Su, Simon Kornblith, Cliff Young, Eric Jang, Sean McPherson, Jonathan Frankle, Carles Gelada, David Ha, Brian Spiering, Samy Bengio, Stephanie Sher, Jonathan Binas, Pete Warden, Sean Mcpherson, Laura Florescu, Jacques Pienaar, Chip Huyen, Raziel Alvarez, Dan Hurt and Kevin Swersky. Thanks for the institutional support and encouragement of Natacha Mainville and Alexander Popper.
}

\clearpage

\bibliography{main}
\bibliographystyle{neurips_2020}
\end{document}